\documentclass[aps,prd,amsmath,amssymb,preprintnumbers]{revtex4}
\pdfoutput=1
\usepackage{multirow}
\usepackage{slashbox}
\usepackage{pict2e} 
\usepackage{graphicx}
\usepackage{color}
\newcommand{\beq}{\begin{eqnarray}}
\newcommand{\eeq}{\end{eqnarray}}

\newcommand{\bmp}{\noindent\begin{minipage}{16cm}}
\newcommand{\emp}{\end{minipage}\vskip 7mm} 

\usepackage{dcolumn}
\usepackage{bm}
\usepackage{bbm}
\usepackage{subfigure}
\usepackage{pxfonts}
\usepackage{booktabs}
\usepackage{epsfig}

\usepackage[ margin=5pt, font=normalsize,labelfont=bf,justification=raggedright]{caption}
\usepackage{youngtab}
\usepackage{slashed}

\definecolor{rossoCP3}{cmyk}{0,.88,.77,.40}

\usepackage{fancyhdr}
\def\lsim{\mathrel{\rlap{\lower4pt\hbox{\hskip1pt$\sim$}}
    \raise1pt\hbox{$<$}}}                
\def\gsim{\mathrel{\rlap{\lower4pt\hbox{\hskip1pt$\sim$}}
    \raise1pt\hbox{$>$}}}                

\begin{document}
\title{\Large  \color{rossoCP3}  W' and Z' limits for Minimal Walking Technicolor}
\author{Jeppe R. Andersen$^{\color{rossoCP3}{\varheartsuit}}$}\email{andersen@cp3-origins.net} 
\author{Tuomas Hapola$^{\color{rossoCP3}{\varheartsuit}}$}\email{hapola@cp3-origins.net} 
\author{Francesco  Sannino$^{\color{rossoCP3}{\varheartsuit}}$}\email{sannino@cp3-origins.net} 
\affiliation{
$^{\color{rossoCP3}{\varheartsuit}}${ \color{rossoCP3}  \rm CP}$^{\color{rossoCP3} \bf 3}${\color{rossoCP3}\rm-Origins} \& the Danish Institute for Advanced Study {\color{rossoCP3} \rm DIAS},\\ 
University of Southern Denmark, Campusvej 55, DK-5230 Odense M, Denmark.
}
\begin{abstract} 
  We interpret the recent data on non-observation of Z'- and W'-bosons, reported by
  CMS, within Minimal Walking Technicolor models and use them to constrain the couplings and spectrum of the theory. We provide the reach for both exclusion and possible observation for the LHC with 5 fb$^{-1}$ at 7 TeV in the centre of mass energy, and 100 fb$^{-1}$ at 13 TeV.  \\
  [.1cm] {\footnotesize \it Preprint: CP$^3$-Origins-2011-16 \&  DIAS-2011-02}
 \end{abstract}

\maketitle
\thispagestyle{fancy}

\section{Introduction}

Recently the CMS \cite{Chatrchyan:2011dx} and ATLAS \cite{Aad:2011fe}
collaborations have reported limits on the masses of possible new spin one
resonances, appearing in several Standard Model extensions. These studies are
performed using 36-40 pb$^{-1}$ of data collected during the year 2010. We interpret these results within Minimal Walking Technicolor (MWT) models, where the new spin zero resonances emerge as composite states of the fundamental degrees of freedom.

The electroweak symmetry can naturally break through
the formation of a chiral condensate, caused by the existence of new strong
dynamics \cite{Weinberg:1979bn,Susskind:1978ms}. Theories under the mechanism
of dynamical electroweak symmetry breaking (DESB) are called
Technicolor. Due to the recent progress
\cite{Sannino:2004qp,Hong:2004td,Gies:2005as,Braun:2005uj,Braun:2006jd,Dietrich:2006cm,Ryttov:2007sr,Ryttov:2007cx,Dietrich:2005jn,Sannino:2008ha,Braun:2009ns,Antipin:2009wr,Jarvinen:2009fe,Mojaza:2010cm,Braun:2010cy,Alanen:2010tg,Pica:2010mt,Pica:2010xq,Frandsen:2010ej,Ryttov:2010iz,Chen:2010er,Ryttov:2010hs,Ryttov:2010jt,Mojaza:2010cm,Mojaza:2011rw,Braun:2010qs,Jarvinen:2010ks} in the
understanding of near conformal (walking) dynamics \cite{Holdom:1981rm,Holdom:1984sk} various phenomenologically viable models have been proposed. Primary examples are: the $SU(2)$ gauge theory with two techniflavors in the adjoint representation,  known as Minimal Walking Technicolor (MWT) \cite{Sannino:2004qp}; the $SU(3)$ theory with
two flavors in the two-index symmetric representation which is called Next to Minimal
Walking Technicolor (NMWT) \cite{Sannino:2004qp} and the $SU(2)$ theory with two techniflavors in the fundamental representation and one techniflavor in the adjoint representation known as Ultra Minimal Technicolor (UMT) \cite{Ryttov:2008xe}. These gauge theories possess remarkable 
properties~\cite{Sannino:2004qp,Dietrich:2005jn,Dietrich:2006cm,Ryttov:2007sr,Ryttov:2007cx} and have the smallest effects on precision data when used for Technicolor~\cite{Sannino:2004qp,Dietrich:2005jn,Foadi:2007ue,Foadi:2007se,Foadi:2008xj,Sannino:2010ca,Sannino:2010fh,DiChiara:2010xb}. We have also shown in \cite{Fukano:2010yv} that the effects of the extensions of the Technicolor models, needed to give masses to the SM fermion, cannot be neglected and lead to models of {\it ideal walking} of which MWT technicolor models are prime examples.  The use of fermions in the fundamental representation of the underlying Technicolor gauge group ensures minimal corrections to the precision data  \cite{Sannino:2010ca,Sannino:2010fh,DiChiara:2010xb}. A simple way to achieve this is to gauge, at most, one doublet of techniquarks, as done for the UMT model. More generally these models are known as the partially gauge Technicolor models and were introduced in \cite{Dietrich:2005jn,Dietrich:2006cm}.   
The theories underlying the models above are being subject to intensive numerical investigations via lattice simulations \cite{Catterall:2007yx, DelDebbio:2008wb,Shamir:2008pb,Deuzeman:2008sc,DelDebbio:2008zf,Catterall:2008qk,DelDebbio:2008tv,DeGrand:2008kx,Hietanen:2008mr, Hietanen:2009az,Deuzeman:2009mh,Hasenfratz:2009ea,DelDebbio:2009fd,Fodor:2009wk,Fodor:2009ar,DeGrand:2009hu,Catterall:2009sb,Bursa:2009we,Bilgici:2009kh,Kogut:2010cz,Hasenfratz:2010fi,DelDebbio:2010hu,DelDebbio:2010hx,Catterall:2010du,Fodor:2011tw}. 
 The important progress of understanding the phase diagram has led to models where the walking dynamics arises with the minimal matter content serving a natural way to go beyond the Standard Model without fundamental scalars. To motivate better the paradigm of complete absence of fundamental scalars, the authors of \cite{Channuie:2011rq} have shown that also a successful description of the inflation can be achieved using strong dynamics featuring solely fermionic matter.

\section{The model}

In order to respect the triumph of the Standard Model, the extensions have to accommodate the following chiral symmetry breaking pattern:
\begin{equation}
SU(2)_{\text{L}}\times SU(2)_{\text{R}}\rightarrow SU(2)_{\text{V}}.
\end{equation}
An example of a walking technicolor model embodying just this pattern for the breaking of the
global flavor symmetry is the Next to Minimal Walking Technicolor model (NMWT).  Any other minimal extension contains the symmetries of NMWT and therefore the common sectors are simultaneously constrained. 
Here the three technipions produced in
the breaking of the symmetry are eaten by the W and Z bosons. The low energy spectrum of the theory is described  by the spin one vector and axial-vector iso-triplets $V^{\pm,0}$, $A^{\pm,0}$ and the lightest iso-singlet scalar state $H$. The scalar state can be naturally as light as the SM Higgs \cite{Sannino:1999qe,Hong:2004td,Sannino:2008ha,Dietrich:2005jn,Dietrich:2006cm,Doff:2008xx,Fabbrichesi:2008ga}.

The effective Lagrangian \footnote{The custodial technicolor limit \cite{Appelquist:1998xf,Appelquist:1999dq,Foadi:2007se} for certain values of the effective Lagrangian parameters reproduces the BESS model (D-BESS)  \cite{Casalbuoni:1985kq,Casalbuoni:1995qt} for which results are available in the literature \cite{Casalbuoni:1997bv, Dominici:1997zh,Battaglia:2002sr}. Following the initial work of \cite{Appelquist:1998xf,Appelquist:1999dq} several effective Lagrangians similar to the one used here have been proposed in the literature \cite{Sekhar Chivukula:2007ic,Lane:2009ct}, however the present approach is the most general one since it includes besides spin one also the spin zero state. Furthermore, we also provide a direct link, via modified WSRs, to the underlying gauge dynamics. Although this link must be taken with the grain of salt, it is not present in other more phenomenological approaches.}, given in the Appendix, contains the following set of parameters:
\begin{itemize}
\item $g$ and $g'$, the $SU(2)_L \times U(1)_Y$ gauge couplings;
\item $\mu$ and $\lambda$, the parameters of the scalar potential;
\item $\tilde{g}$, the strength of the spin one resonances interaction;
\item $m^2$, the $SU(2)_L \times SU(2)_R$ invariant vector-axial mass squared;
\item $r_2, r_3,s$, the couplings between the Higgs and the vector states;
\end{itemize}

It is convenient to express three of the parameters in terms of  $G_{F}$, $M_{Z}$ and $\alpha$, fixed by the experiments. The parameters $r_{2}$, $r_{3}$ and $m$ can be written in terms of the bare axial and vector masses $M_{A}$, $M_{V}$ as follows: 
\begin{align}
m^2 &= \left( M_A^2 + M_V^2 - \tilde{g}^2 v^2 s \right) / 2 \ ,
\\
r_2 &= 2 (M_A^2 - M_V^2) / \tilde{g}^2 v^2 \ ,
\\
r_3 &= 4 M_A^2 \left( 1 \pm \sqrt{1 - \tilde{g}^2 S / 8 \pi} \right) / \tilde{g}^2 v^2 \ .
\end{align}
S is the S-parameter obtained using the zeroth Weinberg Sum Rule (0th WSR). The effective Lagrangian and the features of the theory were first introduced in \cite{Appelquist:1999dq,Foadi:2007ue} and efficiently summarized in \cite{Andersen:2011yj}. 
Furthermore,  $M_{V}$ and S can be related using the 1st WSR and S is set to 0.3, naively estimated from the underlying Technicolor theory  \cite{Appelquist:1999dq,Foadi:2007ue}. 
Thus we are left with the $M_{A}$, $\tilde{g}$, $M_{H}$ and $s$ as parameters we can vary at the effective Lagrangian level. First principle lattice simulations should be able to determine these parameters in the near future. The walking dynamics nature is associated to the modification of the second WSR according to the discovery made in \cite{Appelquist:1998xf}.  The following caveats apply when matching the low energy effective theory to the underlying gauge theory using the modified WSRs \cite{Appelquist:1998xf} as clearly stated in the original paper: One assumes the narrow width approximation;  the spectrum is approximated at very low energies by a few resonances while the walking dynamics is modeled using dynamical mass for the technifermions respecting the Schwinger-Dyson scaling behavior. 

The new vector and axial-vector states mix with the SM gauge eigenstates yielding the ordinary SM bosons and two triplets of heavy mesons,  $R_{1}^{\pm,0}$ and $R_{2}^{\pm,0}$, as mass eigenstates. The couplings of the heavy mesons to the SM particles are induced by the mixing. Momentous for this study is how the heavy vectors couple to the fermions. In the region of parameter space where $R_{1}$ is mainly an axial-vector and $R_{2}$ mainly a vector sate, the dependence of the couplings to the SM fermions as a function of $\tilde{g}$ is very roughly 
\begin{equation}
g_{R_{1,2}f\bar{f}}\sim \frac{g^2}{\tilde{g}}
\label{gt}
\end{equation}
where $g$ is the electroweak gauge coupling. The full coupling constant is also a function of $M_{A}$, but this dependence is very weak.

\section{Slicing the Parameter Space}

To constrain the parameter space of the model we use the recent CMS results
\cite{Chatrchyan:2011dx}, which report limits for a $W^{\prime}$ boson
decaying to a muon and a neutrino at  $\sqrt{s} = 7$ TeV in the mass range
600 - 2000 GeV for the resonance. ATLAS results are comparable, but the CMS
limits are slightly more stringent in the lower mass region, as shown in figure 5 of  \cite{Aad:2011fe}. Also, the existing data for the dilepton final state can be used to perform this analysis, but currently the limits are weaker than for the lepton plus missing energy final state.

The relevant calculations are performed using MadGraph \cite{Alwall:2007st}, using the CTEQ6L parton distribution functions \cite{Pumplin:2002vw}, 
and the implementation for the NMWT
model \cite{FRweb}. \cite{Chatrchyan:2011dx} reports experimental limits for the muon
channel together with a combined analysis with the electron channel
\cite{Khachatryan:2010fa}. Slightly different cuts are used for the electron
and muon channel. We simplify the analysis slightly, using the most limiting
combination of cuts for the theoretical predictions (this leads to
insignificant differences). Because of the missing energy in the final state, the
invariant mass of the resonance cannot be reconstructed, and the following transverse mass variable is utilized 
\begin{equation}
M_{T}=\sqrt{2\cdot p_{T}E_{T}^{\text{miss}}\cdot(1-\cos\Delta\phi_{\mu,\nu})}.
\end{equation}
When considering only two-to-two processes with zero mass final state
particles, the angle between the lepton and neutrino is fixed and this
definition of the transverse mass reduces to $M_{T}=2\cdot p_{T}$. In the
experimental analysis, the cut on the transverse mass is adjusted in bins of
the mass of the sought after resonance.  The minimum transverse mass cuts and the physical masses of the vector mesons for three different values of $\tilde{g}$ are given in Table \ref{tab}. The decay widths of the vector mesons are given in Table \ref{tab:width}. 
\begin{table}[bp]
\caption{Minimum $M_{T}$ values for different values of the $M_{A}$.The last
  six columns give the physical masses of the $M_{R_{1,2}}$ as a function of
  $M_{A}$, for three different values of the $\tilde{g}$.}
\begin{center}
\begin{tabular}{cccccccc}
\hline
& & \multicolumn{2}{c}{$\tilde{g}=2$}&\multicolumn{2}{c}{$\tilde{g}=4$}&\multicolumn{2}{c}{$\tilde{g}=6$} \\
\cline{3-8}
$M_{A}$ (GeV)&$M_{T}$ (GeV)&$M_{R_{1}}$ (GeV) &$M_{R_{2}}$ (GeV)&$M_{R_{1}}$ (GeV) &$M_{R_{2}}$  (GeV) &$M_{R_{1}}$ (GeV) &$M_{R_{2}}$ (GeV)  \\
\hline
\hline
600 & 400 		&612	&704	&603 				&887			& 601	& 1142\\
700 & 500 		&714	&794	&704 				&945 				& 701	& 1174\\
800 & 500 		&814	&888	&804 				&1008 				& 801	& 1210\\
900 & 500 		&913	&983	&905 				&1075 				& 902	& 1250\\
1000 & 530 		&1012	&1081	&1005 				&1145 				& 1002	& 1293\\
1100 & 590 		&1111	&1180	&1106 				&1218 				& 1102	& 1339\\
1200 & 650 		&1209	&1280	&1206 				&1294 				& 1202	& 1387\\
1300 & 675 		&1307	&1382	&1306 				&1371 				& 1302	& 1438\\
1400 & 675 		&1404	&1484	&1406 				&1450 				& 1402	& 1491\\
1500 & 680 		&1502	&1586	&1505 				&1532 				& 1502	& 1546\\
2000 & 690 		&1991	&2102	&1940 				&2013 				& 1842	& 2003\\
\hline
\end{tabular}
\end{center}
\label{tab}
\end{table}
\begin{table}[bp]
\caption{Spin one widths for different values of $M_A$ and $\tilde{g}$. Note that the increase in the widths for certain ranges of these couplings is due to the opening of new channels as explained in \cite{Andersen:2011yj}}
\begin{center}
\begin{tabular}{ccccccc}
\hline
&  \multicolumn{2}{c}{$\tilde{g}=2$}&\multicolumn{2}{c}{$\tilde{g}=4$}&\multicolumn{2}{c}{$\tilde{g}=6$} \\
\cline{2-7}
$M_{A}$ (GeV)&$\Gamma_{R_{1}}$ (GeV) &$\Gamma_{R_{2}}$ (GeV)&$\Gamma_{R_{1}}$ (GeV) &$\Gamma_{R_{2}}$  (GeV) &$\Gamma_{R_{1}}$ (GeV) &$\Gamma_{R_{2}}$ (GeV)  \\
\hline
\hline
600& 0.678 & 1.89 & 0.467 & 57.0 & 1.88 & 258 \\
700& 0.675 & 2.24 & 0.754 & 36.7 & 2.96 & 187 \\
800& 0.669 & 2.71 & 1.20 & 23.0 & 4.45 & 133 \\
900& 0.699 & 3.25 & 1.86 & 14.2 & 6.50 & 93.1 \\
1000& 0.809 & 3.84 & 2.83 & 9.23 & 9.27 & 64.5 \\
1100& 1.04 & 4.50 & 4.21 & 7.27 & 12.9 & 45.8 \\
1200& 1.44 & 5.25 & 6.11 & 7.83 & 17.7 & 35.6 \\
1300& 2.04 & 6.14 & 8.69 & 10.3 & 23.9 & 32.4 \\
1400& 2.89 & 7.22 & 12.1 & 13.4 & 31.7 & 35.1 \\
1500& 4.02 & 8.55 & 16.6 & 17.4 & 41.5 & 41.9 \\
2000& 16.2 & 21.9 & 57.1 & 61.6 & 100 & 138\\
\hline
\end{tabular}
\end{center}
\label{tab:width}
\end{table}
In addition to the transverse mass cut, the lepton acceptance $|\eta|<2.1$ is
used. The resulting cross section is then compared with the limits reported
in the experimental analyses.

\begin{figure}[tbp]
\begin{center}
 \includegraphics[width=0.75\textwidth]{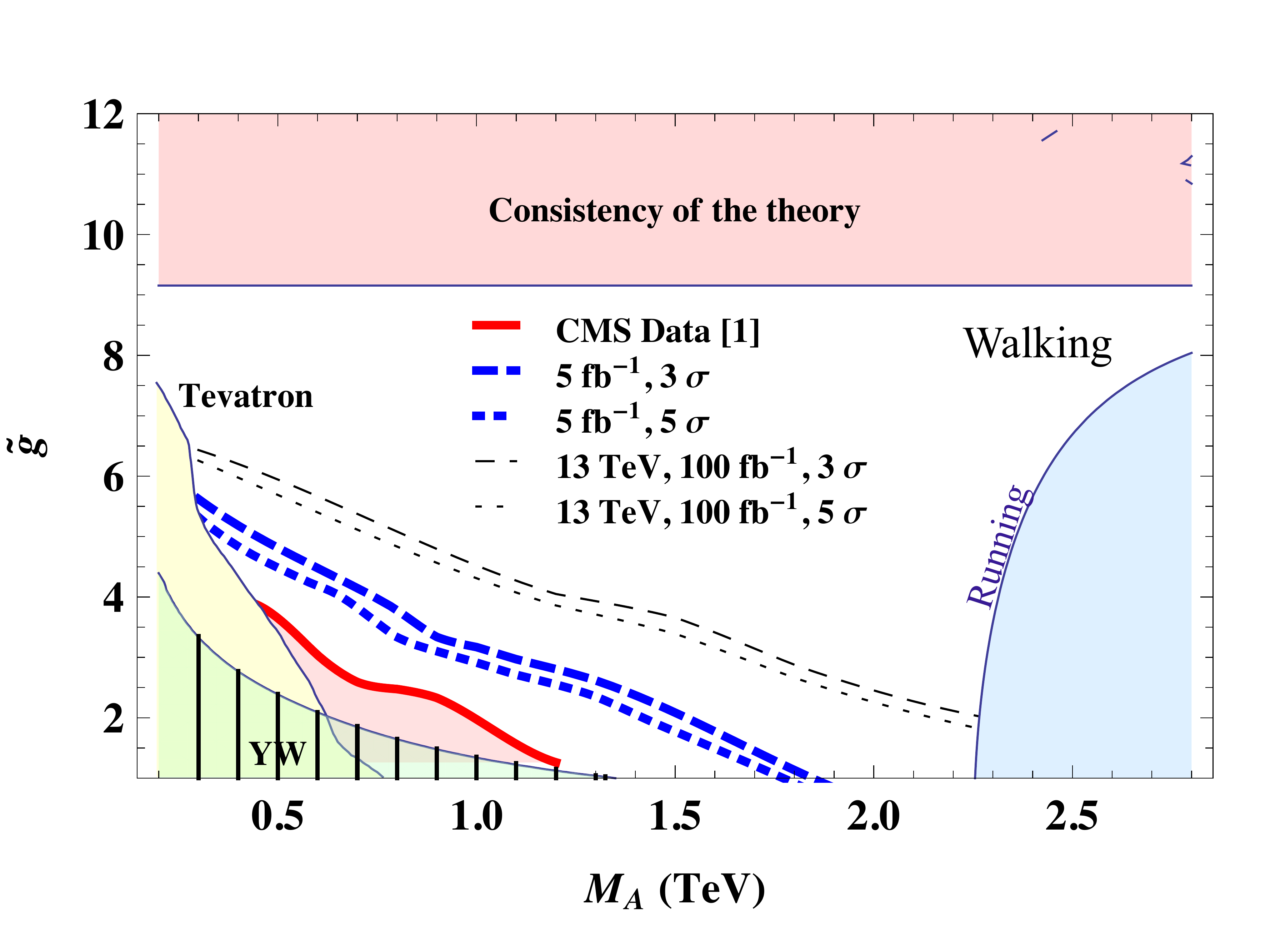}
\caption{Bounds in the ($M_{A}$, $\tilde{g}$) plane of the NMWT parameter space: (i) CDF direct searches of the neutral spin one resonance excludes the uniformly shaded area in the left, with $M_{H} = 200$ GeV and $s=0$. (ii) The 95 \% confidence level measurement of the electroweak precision parameters W and Y excludes the striped area in the left corner. (iii) Imposing the modified WRS's excludes the uniformly shaded area in the right corner. (iv) The horizontal stripe is excluded imposing reality of the axial and axial-vector decay constants. (v) The area below the thick uniform line is excluded by the CMS data  \cite{Chatrchyan:2011dx}. (vi) Dashed and dotted lines are expected exclusions using different values of the integrated luminosity and center of mass energy.}
\label{pspace}
\end{center}
\end{figure}

Exploring the signal from the process $pp \to R_{1,2} \to l\nu$ we are able to limit the possible values for the parameters $M_{A}$ and $\tilde{g}$. The theoretical limits as well as the limits from the Tevatron are described in \cite{Belyaev:2008yj}. The $M_{A}$, $\tilde{g}$ plane of the parameter space is presented in Fig.~\ref{pspace} for $M_{H}=200 GeV$ and $s=0$. 

The uniformly shaded region on the left is excluded by the CDF searches of the
resonance in the the $p\bar{p}\to e^+e^-$ process. The striped region in the
lower left corner is excluded by the measurements of the electroweak $W$ and
$Y$ parameters \cite{Barbieri:2004qk}  adapted for models of MWT in \cite{Foadi:2007se}. Avoiding imaginary decay constants for the vector and
axial-vector sets an upper bound for the $\tilde{g}$, i.e. excludes the
uniformly shaded in the upper part of the figure. The near conformal
(walking) dynamics modifies the WRS's, compared to a running case like
QCD, as explained above \cite{Appelquist:1998xf}. Imposing these modified sum rules excludes the lower right corner of the
parameter space. The CDF exclusion limit is sensitive, indirectly, to the mass of the composite Higgs and the coupling
$s$ via properties of the new heavy spin one states. However, the edge of the excluded area varies only very weakly as a function
of $s$ and $M_H$. The CMS search imposes a 95 \% CL exclusion bound described with the
thick solid (red) line. The thick dashed and dotted lines (blue) are three and five sigma
exclusion limits for  7~TeV and 5~$fb^{-1}$. The thin dotted and dashed lines
describe the reach of the LHC with 100~$fb^{-1}$ at 13~TeV. The three and
five sigma exclusion limits are calculated using poisson distribution,
following \cite{Aad:2009wy}. Due to the effective description, we have not employed the $K$-factors when calculating the exclusion limits. In Table~\ref{tab2} we report the explicit values of the signal cross section for given points in the parameter space. 


Comparing the three sets of lines for the LHC, the increase in the horizontal
direction follows roughly the increase in luminosity. The small role of the
center of mass energy can be understood by exploring the behavior of the
cross section as a function of the center of mass energy and comparing it
with the scaling with $\tilde{g}$, obtained from equation \eqref{gt}. The
Tevatron exclusion line behaves completely differently compared to the lines
calculated for the LHC. This follows from the two distinct feature between
the machines: In the relevant kinematic, the pdf's are highly gluon dominated
at the LHC energies, and the LHC is a proton-proton collider, which further
suppresses the portion of the anti-quarks inside the colliding protons compared
to the Tevatron.

\section{Conclusions}

We have interpreted experimental bounds from the LHC on new resonances in the
$e,\nu_e$ and $\mu, \nu_\mu$-channel within the sector common to all the Minimal Walking Technicolor models. We demonstrated that the data from the LHC are already excluding new
regions of the parameter space of strongly coupled extensions of the SM.  However much of the parameter space is still allowed,
and will be probed by the data arriving in the near future.
\begin{table}[h]
\caption{Signal cross sections for given points in the parameter space.}
\begin{center}
\begin{tabular}{l|ccccc}
\hline
~~~7 TeV& \multicolumn{4}{c}{$\sigma$ (fb)} \\ \hline
\backslashbox{~$M_{A}$}{$\tilde{g}$}&2&2.5&3&3.5&4 \\ \hline\hline
600 (GeV)& - &123& 70&39 & 19      \\ 
800 & 100 &50& 17 &7.0& 3.0    \\ 
1000 & 37 &17& 6.6 &2.1 &0.75\\
\hline
\end{tabular}
\hspace{50pt}
\begin{tabular}{l|ccccc}
\hline
~~13 TeV& \multicolumn{5}{c}{$\sigma$ (fb)} \\ \hline
\backslashbox{~$M_{A}$}{$\tilde{g}$}&3&4&5&6&7 \\ \hline\hline
600 (GeV) &- &-& 9.4 & 1.4 & 0.17     \\ 
1500 & 3.4 & 0.50 & 0.10&0.027&0.0085     \\ 
2100 & 0.20 & 0.033 & 0.0089&-&- \\
\hline
\end{tabular}
\end{center}
\label{tab2}
\end{table}

\newpage
\appendix

\section{Effective Lagrangian}

The composite spin-1 and spin-0 states and their interaction with the SM fields are described via the following effective Lagrangian  we developed in \cite{Foadi:2007ue,Appelquist:1999dq}: 

\begin{eqnarray}
{\cal L}_{\rm boson}&=&-\frac{1}{2}{\rm Tr}\left[\widetilde{W}_{\mu\nu}\widetilde{W}^{\mu\nu}\right]
-\frac{1}{4}\widetilde{B}_{\mu\nu}\widetilde{B}^{\mu\nu}
-\frac{1}{2}{\rm Tr}\left[F_{{\rm L}\mu\nu} F_{\rm L}^{\mu\nu}+F_{{\rm R}\mu\nu} F_{\rm R}^{\mu\nu}\right] \nonumber \\
&+& m^2\ {\rm Tr}\left[C_{{\rm L}\mu}^2+C_{{\rm R}\mu}^2\right]
+\frac{1}{2}{\rm Tr}\left[D_\mu M D^\mu M^\dagger\right]
-\tilde{g^2}\ r_2\ {\rm Tr}\left[C_{{\rm L}\mu} M C_{\rm R}^\mu M^\dagger\right] \nonumber \\
&-&\frac{i\ \tilde{g}\ r_3}{4}{\rm Tr}\left[C_{{\rm L}\mu}\left(M D^\mu M^\dagger-D^\mu M M^\dagger\right)
+ C_{{\rm R}\mu}\left(M^\dagger D^\mu M-D^\mu M^\dagger M\right) \right] \nonumber \\
&+&\frac{\tilde{g}^2 s}{4} {\rm Tr}\left[C_{{\rm L}\mu}^2+C_{{\rm R}\mu}^2\right] {\rm Tr}\left[M M^\dagger\right]
+\frac{\mu^2}{2} {\rm Tr}\left[M M^\dagger\right]-\frac{\lambda}{4}{\rm Tr}\left[M M^\dagger\right]^2
\label{eq:boson}
\end{eqnarray}
where $\widetilde{W}_{\mu\nu}$ and $\widetilde{B}_{\mu\nu}$ are the ordinary electroweak field strength tensors, $F_{{\rm L/R}\mu\nu}$ are the field strength tensors associated to the vector meson fields $A_{\rm L/R\mu}$~\footnote{In Ref.~\cite{Foadi:2007ue}, where the chiral symmetry is SU(4), there is an additional term whose coefficient is labeled $r_1$. With an SU($N$)$\times$SU($N$) chiral symmetry this term is just identical to the $s$ term.}, and the $C_{{\rm L}\mu}$ and $C_{{\rm R}\mu}$ fields are
\begin{eqnarray}
C_{{\rm L}\mu}\equiv A_{{\rm L}\mu}-\frac{g}{\tilde{g}}\widetilde{W_\mu}\ , \quad
C_{{\rm R}\mu}\equiv A_{{\rm R}\mu}-\frac{g^\prime}{\tilde{g}}\widetilde{B_\mu}\ .
\end{eqnarray}
The 2$\times$2 matrix $M$ is
\begin{eqnarray}
M=\frac{1}{\sqrt{2}}\left[v+H+2\ i\ \pi^a\ T^a\right]\ ,\quad\quad  a=1,2,3
\end{eqnarray}
where $\pi^a$ are the Goldstone bosons produced in the chiral symmetry breaking, $v=\mu/\sqrt{\lambda}$ is the corresponding VEV, $H$ is the composite Higgs, and $T^a=\sigma^a/2$, where $\sigma^a$ are the Pauli matrices. The covariant derivative is
\begin{eqnarray}
D_\mu M&=&\partial_\mu M -i\ g\ \widetilde{W}_\mu^a\ T^a M + i\ g^\prime \ M\ \widetilde{B}_\mu\ T^3\ . 
\end{eqnarray}
When $M$ acquires its VEV, the Lagrangian of Eq.~(\ref{eq:boson}) contains mixing matrices for the spin one fields. 


\end{document}